\documentclass[twoside,reqno]{H}
\usepackage{epsfig,cite,colordvi}
\usepackage{graphicx}
\usepackage{graphics}
\usepackage{url}
\usepackage{amssymb,amsmath,amscd,epsf}
\usepackage{times}
\usepackage{makeidx}
\usepackage[english]{babel}
\def\r{\mbox{\boldmath $r$}}

\def\q{\mbox{\boldmath $q$}}
\def\diff{{\rm\,d}}                    
\def\mcv{\mbox{$\mathcal{V}$}}

\newcommand{\eep} { $(e,e^{\,\prime}p)$ } 
\def\ee{\mbox{$\left(e,e^{\prime}\right)$\ }}
\newcommand{\bint}{\mskip .5mu \int \mskip-18mu} 

\begin{document}

\title{The Relativistic Green's Function Model for Quasielastic Neutrino-Nucleus Scattering}

\runningheads{Relativistic Descriptions of Final-State Interactions in Quasielastic
Electron and Neutrino-Nucleus Scattering}{C. Giusti, A. Meucci}

\begin{start}

\author{C. Giusti}{1}, \coauthor{A. Meucci}{1},

\index{Giusti, Carlotta}
\index{Meucci, Andrea}

\address{Dipartimento di Fisica, 
Universit\`{a} degli Studi di Pavia and \\
INFN, Sezione di Pavia, via Bassi 6 I-27100 Pavia, Italy}{1}

\begin{Abstract}
A model based on the relativistic impulse approximation (RIA) for quasielastic 
(QE) lepton-nucleus scattering is presented. The effects of the final-state
interactions (FSI) between the emitted nucleon and the residual nucleus are 
described by the relativistic Green's function (RGF) model where FSI are treated
consistently with the exclusive scattering and using the same complex optical 
potential. The results of the model are compared with the results of different 
descriptions of FSI and with available data for neutrino-nucleus scattering.
\end{Abstract}
\end{start}

\section{Introduction}
A deep understanding of neutrino-nucleus cross sections is very important for 
the determination of neutrino oscillation parameters. Reliable models are 
required where all nuclear effects are well under control. Within the
QE kinematic domain, the nuclear response to an electroweak probe is dominated
by one-nucleon processes, where the scattering occurs only with the nucleon
which is emitted and the remaining nucleons of the target behave as spectators.
The process can adequately be described in the RIA by the sum of
incoherent processes involving only one nucleon scattering. The components 
of the hadron tensor are then obtained from the 
sum, over all the single-particle (s.p.) shell-model states, of the squared 
absolute value  of the transition matrix elements of the single-nucleon 
current. A reliable description of FSI between the emitted nucleon and the
residual nucleus is a crucial ingredient for a proper description of
data. The relevance of FSI has been clearly established  for the exclusive 
\eep reaction, where FSI are described in the nonrelativistic or relativistic 
distorted-wave impulse approximation (DWIA or RDWIA) using a complex optical 
potentials (OP) whose imaginary part gives an absorption that reduces the
calculated  cross section~\cite{rep,book,ud93,meucci01,meucci01a,epja,Giusti11}. 
In the optical model the imaginary part 
accounts for the fact that in the elastic nucleon-nucleus scattering, if other 
channels are open besides the elastic one, part of the incident 
flux is lost in the elastically scattered beam and goes to the inelastic 
channels which are open. In the exclusive \eep reaction, where the emitted 
proton is detected in coincidence with the scattered electron and the residual 
nucleus is left in a specific discrete eigenstate, only the selected channel 
contributes and it is correct to account for the flux lost in the considered 
channel.  In the inclusive scattering, where only the scattered lepton is
detected and the final nuclear state is not determined, all elastic and 
inelastic channels contribute, the flux lost in a channel must be recovered 
in the other channels, and in the sum over all the channels the flux can be 
redistributed but must be conserved. A model based on the DWIA, where the cross section is 
obtained from the sum over all integrated one-nucleon knockout channels and FSI 
are described by a complex OP with an absorptive imaginary part, does not 
conserve the flux and it is conceptually wrong. 

Different approaches have been adopted
within the RIA 
to describe FSI in the inclusive QE lepton-nucleus scattering. 
In the relativistic plane-wave impulse approximation (RPWIA) the plane-wawe
approximation is assumed for the emitted nucleon wave function and FSI 
are simply neglected. In some RIA calculations FSI are incorporated in the
emitted nucleon states by using real potentials, either retaining only the real
part of the relativistic OP (rROP) or using the same relativistic mean-field
potential considered in the description of the initial nucleon state
(RMF)~\cite{Chiara03,cab,cab1,confee,confcc}.  
 
In the RGF model FSI are described in the inclusive scattering consistently with
the exclusive scattering by the same complex ROP, but the imaginary part is used
in the two cases in a different way and in the inclusive scattering the flux, 
although redistributed in all the channels, is 
conserved~\cite{eenr,ee,eesym,cc,eea,acta,acta1,confee,confcc,compmini,prd,prd1,eeex,prd2,compnc}.
The Green's function model was originally developed within a
nonrelativistic and then a relativistic framework for the
inclusive electron scattering. But for some differences and complications due 
to the Dirac matrix structure, the formalism follows within both frameworks 
the same steps and approximations. Relativity is, however, important at all 
energies, in particular at high energies, and in the energy regime of many 
neutrino experiments a relativistic model is required, where not only 
relativistic kinematics is considered, but nuclear dynamics and currents 
operators are described within a relativistic framework. Therefore only the 
relativistic RGF model has been extended to neutrino-nucleus scattering. 
 
The RGF model and its results are reviewed in 
this contribution. 

\section{The Green's Function Model }

In the one-boson exchange approximation the cross section for QE lepton-nucleus
scattering is obtained from the contraction between the lepton tensor, which
in the plane-wave approximation for the 
lepton wave functions depends only on the lepton kinematics,  and the hadron 
tensor $W^{\mu\nu}$, whose components 
are given by 
products of the matrix elements of the nuclear current  
$J^{\mu}$ between the initial and final nuclear states, \ie,
\begin{eqnarray}
W^{\mu\nu}(\omega,q) &=& \overline{\sum_i}\bint\sum_f \, \langle 
\Psi_f\mid J^{\mu}(\q) \mid \Psi_i\rangle \, 
\langle \Psi_i \mid J^{\nu\dagger}(\q)\mid \Psi_f\rangle \, \nonumber \\ &\times&
\delta(E_i+\omega-E_f),
\label{eq.wmn}
\end{eqnarray}
where $\omega$ and $\q$ are the energy and momentum transfer, respectively.

For the inclusive scattering the
diagonal components of the hadron tensor can equivalently be expressed as
\begin{equation}
W^{\mu\mu}(\omega,q) = -\frac{1}{\pi} {\rm{Im}} \langle \Psi_i
\mid J^{\mu\dagger}(\q) G(E_{{f}}) J^{\mu}(\q) \mid \Psi_i \rangle \
,\label{eq.hadrten}
\end{equation}
where $E_{{f}} = E_i +\omega$ and $G(E_{{f}})$ is the Green's
function, the full A-body propagator, which is related to the many body 
nuclear Hamiltonian.
A similar but more cumbersome expression is obtained for the
non-diagonal components~\cite{cc}. 

The A-body Green's function in Eq.~(\ref{eq.hadrten})
defies a practical evaluation. Some approximations are required to 
reduce the problem to a tractable form. With suitable approximations, which 
are basically related to the IA, the components of hadron tensor can be 
written in terms of the s.p. optical-model Green's  function. 
This result has been derived by arguments based on multiple scattering 
theory \cite{hori} or by means of projection operators 
techniques~\cite{eenr,ee,eesym,cc}. In the latter framework, the 
matrix element in Eq. (\ref{eq.hadrten}) is decomposed into the sum

\begin{equation}
{\rm{Im}} \langle \Psi_i \mid J^{\mu\dagger} GJ^{\mu} 
\mid \Psi_i \rangle \, \sim A \sum_{n} {\rm{Im}} \langle \Psi_i \mid j^{\mu\dagger} 
G_nj^{\mu}  \mid \Psi_i \rangle, \,  \label{eq.gf1}
\end{equation}
where
\begin{equation}
G_n = P_n G P_n, \,\,\,\,\, P_n = \int \diff \r \mid \r ; n \rangle 
\langle n;\r \mid  \,\label{eq.pn}
\end{equation}
is the projection of the full Green's function onto the channel subspace spanned
by the orthonormalized set of states $\mid \r ; n \rangle$, corresponding to a
nucleon at the point $\r$ and to the residual nucleus in the state  
$\mid n \rangle$. $G_n$ is the the Green's function associated with the OP
Hamiltonian which describes the elastic scattering of a nucleon by the
$(A-1)$-nucleus in the state $\mid n \rangle$. The sum over $n$ includes discrete
eigenstates and resonances embedded in the continuum~\cite{eenr,ee}.

The approximation in Eq.~(\ref{eq.gf1}) has been derived neglecting non-diagonal
terms $P_n G P_m$  and retaining only the one-body part $j^\mu$ of the current
operator. It is basically a s.p.approach where one assumes that $j^\mu$
connects the initial state $\mid \Psi_i \rangle$ only with states in the channel
subspace spanned by the vectors $\mid \r ; n \rangle$, \ie, only with states
asymptotically corresponding to single-nucleon knockout. However, not only these
states, but all the allowed final states are included in the
inclusive response, as $G_n$ in Eq.~(\ref{eq.pn}) contains the full 
propagator $G$: all the allowed
final states are taken into account by the OP, in particular by its imaginary
part.  

The complexity of an explicit calculation of $G_n$ can be avoided by means of
its spectral representation, which is based on a biorthogonal expansion in 
terms of the eigenfunctions of the non-Hermitian OP and of its 
Hermitian conjugate~\cite{eenr,ee}. In the s.p. representation one obtains
\begin{eqnarray}
&  & W^{\mu\mu}(\omega,q) )  =  \sum_n \Bigg[ {\rm{Re}} \ T_n^{\mu\mu}
(E_{{f}}-\varepsilon_n, E_{{f}}-\varepsilon_n)  \nonumber \\ 
& - & \frac{1}{\pi} \mathcal{P}  \int_M^{\infty} \diff \mathcal{E} 
\frac{1}{E_{{f}}-\varepsilon_n-\mathcal{E}} 
\ {\rm{Im}} \ T_n^{\mu\mu}
(\mathcal{E},E_{{f}}-\varepsilon_n) \Bigg] \ , \label{eq.finale}
\end{eqnarray}
where $\mathcal{P}$ denotes the principal value of the integral, $n$ is the 
eigenstate of the residual nucleus with energy 
$\varepsilon_n$, and
\begin{eqnarray}
T_n^{\mu\mu}(\mathcal{E} ,E) &=& \lambda_n  \langle \varphi_n
\mid j^{\mu\dagger}(\q) \sqrt{1-\mcv'(E)}
\mid\tilde{\chi}_{\mathcal{E}}^{(-)}(E)\rangle \nonumber \\
&~ & \times  \langle\chi_{\mathcal{E}}^{(-)}(E)\mid  \sqrt{1-\mcv'(E)} j^{\mu}
(\q)\mid \varphi_n \rangle  \ . \label{eq.tprac}
\end{eqnarray}
In Eq.~(\ref{eq.tprac}) $\tilde{\chi}_{\mathcal{E}}^{(-)}$ and 
$\chi_{\mathcal{E}}^{(-)}$ are eigenfunctions, belonging to the eigenvalue
$\mathcal{E}$,  of the s.p. OP and of its hermitian conjugate, $\varphi_n$ is 
the overlap between $\mid \Psi_ i \rangle$ and $\mid n\rangle$, \ie, a s.p. 
bound state, and the spectroscopic factor 
$\lambda_n$ is the norm of the overlap function. 
The factor $\sqrt{1-\mcv'(E)}$, where 
$\mcv'(E)$ is the energy derivative of the OP, accounts for interference 
effects between different channels and justifies the replacement in the 
calculations of the Feshbach OP $\mcv$ by the 
local phenomenological OP

Disregarding the square root correction, the matrix elements in 
Eq.~(\ref{eq.tprac}) are of the same type as the DWIA ones 
of the exclusive \eep  reaction, but both 
eigenfunctions $\tilde{\chi}_{\mathcal{E}}^{(-)}$ and 
$\chi_{\mathcal{E}}^{(-)}$ of $\mcv(E)$ and of $\mcv^{\dagger}(E)$  are 
considered.  In the exclusive scattering the imaginary part of the OP accounts for the flux lost in the 
channel $n$  towards the channels different from $n$, which are not included 
in the exclusive process. In the inclusive response, where  all the channels are
included, this loss is compensated by a corresponding 
gain of flux due to the flux lost, towards the channel $n$, in the other final 
states asymptotically originated by the channels different from $n$. 
This compensation is performed by the first matrix element in the right hand 
side of Eq.~(\ref{eq.tprac}),  which involves the eigenfunction 
$\tilde{\chi}_{\mathcal{E}}^{(-)}$ of the Hermitian conjugate OP, where the 
imaginary part has an opposite sign and has the effect of increasing the 
strength. Therefore, in the RGF model the imaginary part of the OP 
redistributes the flux lost in a channel in the other channels, and 
in the sum over $n$ the total flux is conserved.  
The RGF model allows to recover the contribution of non-elastic channels starting 
from the complex OP that describes elastic nucleon-nucleus 
scattering data and provides a consistent treatment of FSI in the exclusive and in 
the inclusive scattering.

\section{Results}

The RGF model has been applied to the inclusive QE electron scattering and  
charged-current QE (CCQE) neutrino-nucleus scattering. Some numerical results
are presented and discussed in this Section.

In all the calculations the bound nucleon states 
are self-consistent Dirac-Hartree solutions derived within a relativistic
mean-field approach~\cite{Serot1986}. Different parameterizations  
for the ROP have been used: the energy-dependent and A-dependent EDAD1~\cite{chc} and the more recent 
democratic (DEM)~\cite{chc1} parameterizations, which are global parameterizations  
obtained through a fit to elastic proton-scattering data over a wide range of 
nuclei, and the energy-dependent but A-independent EDAI complex phenomenological 
potential~\cite{chc}. While EDAD1 and DEM depend on the atomic 
number $A$,  EDAI  is constructed  only to fit elastic proton-scattering on a 
specific nucleus, for instance on $^{12}$C, which is of particular interest 
for recent neutrino experiments. 
 
\begin{figure}[t]
\includegraphics[scale=.28]{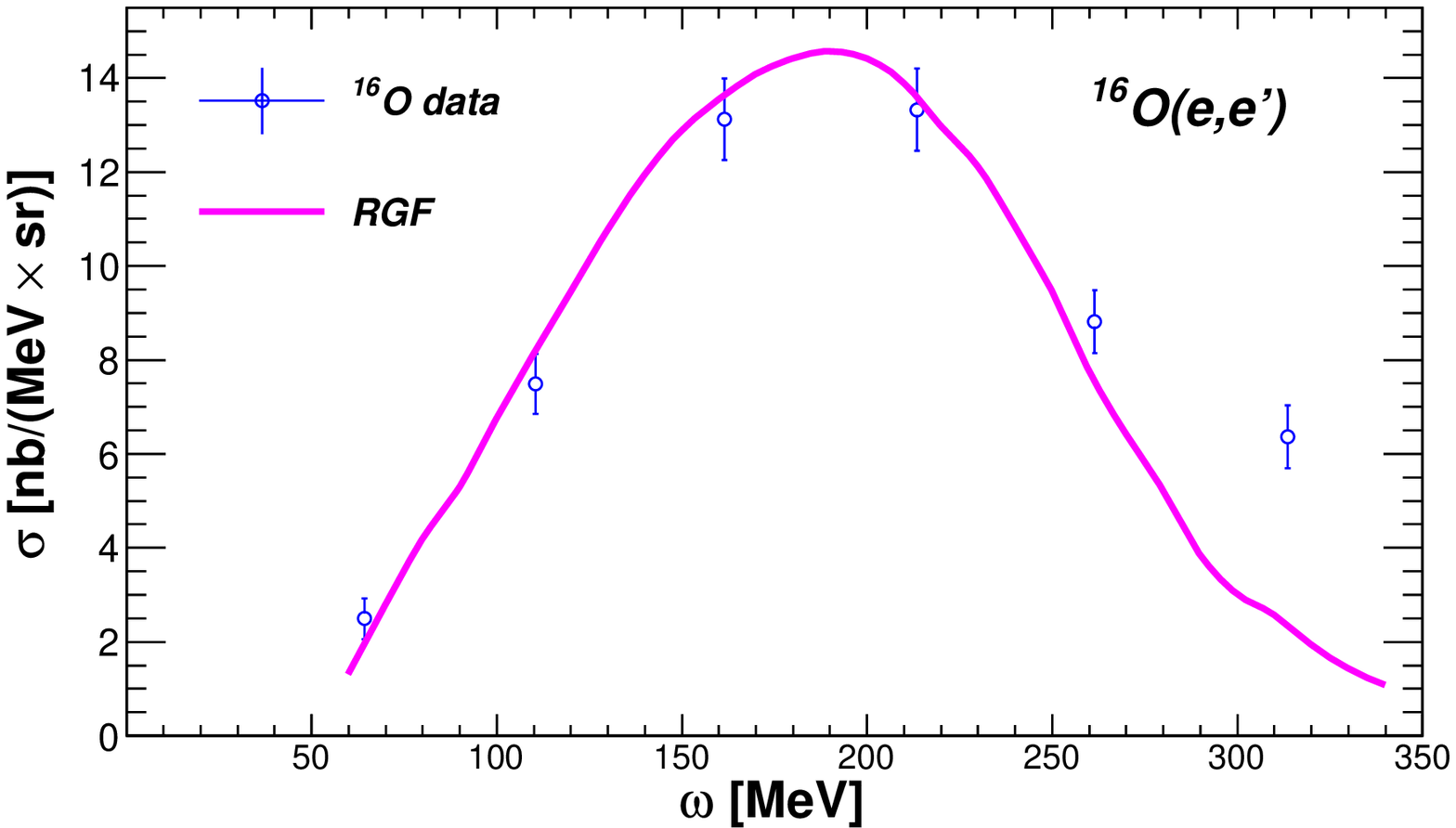}
\includegraphics[scale=.28]{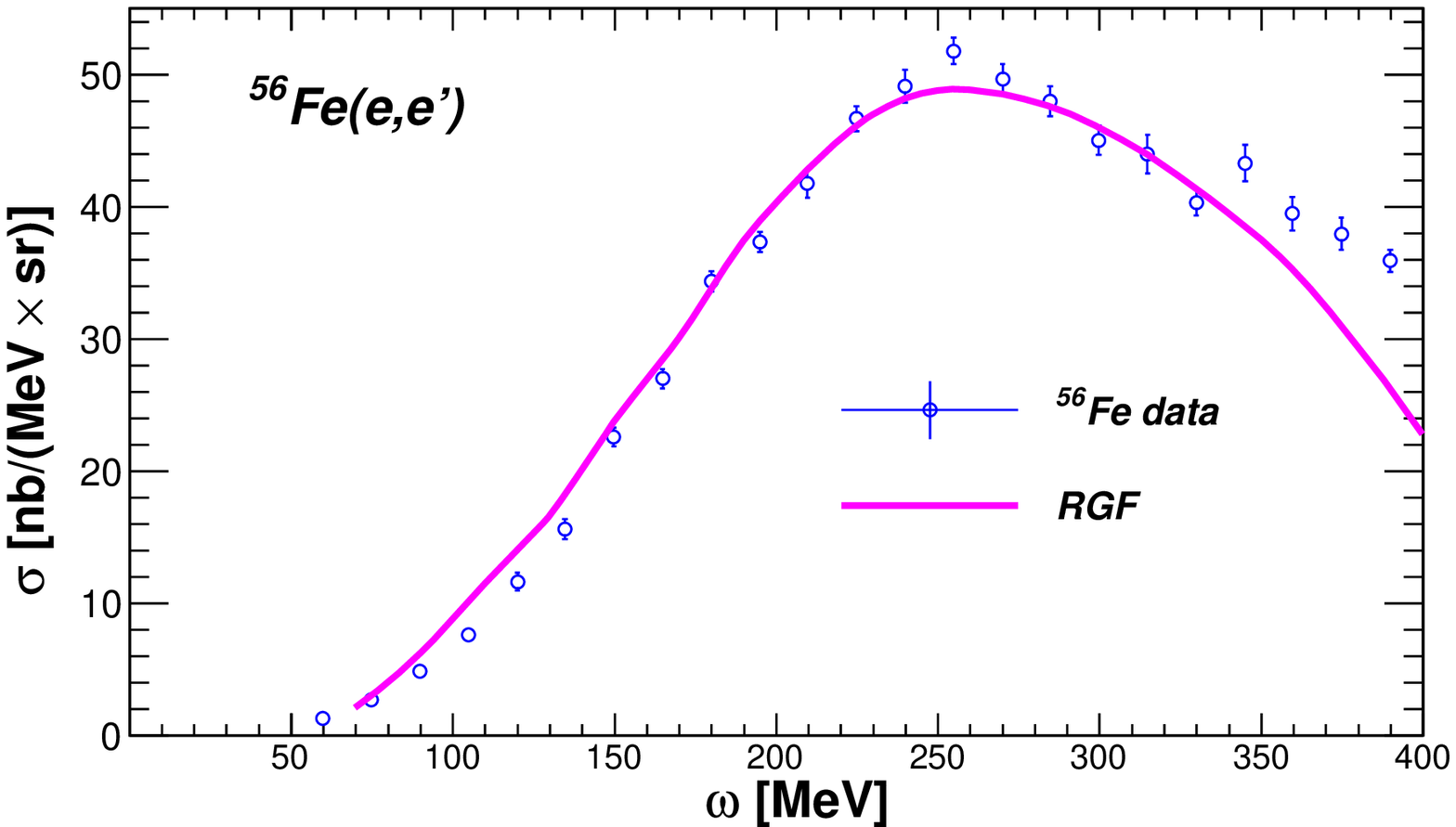}
\caption{
Differential cross sections of the reactions
$^{16}$O$(e,e^{\prime})$  (beam energy  
$\varepsilon = 1080$ MeV and scattering angle $\vartheta = 32^{\mathrm{o}}$)
 and	$^{56}$Fe$(e,e^{\prime})$  ($\varepsilon = 2020$ MeV and 
$\vartheta = 20^{\mathrm{o}}$) calculated in the RGF-DEM. 
Experimental data from \cite{Anghinolfi} ($^{16}$O) 
and \cite{fe56} ($^{56}$Fe).}\label{fexpgf}
	\end{figure}
The RGF model is in many cases able to give a reasonable description 
of the experimental $\ee$ cross sections in the QE region, in particular in
kinematic situations where the longitudinal response gives the main
contribution. A numerical example is shown in Figure~\ref{fexpgf}, where the RGF 
results are compared with the experimental $\ee$ cross sections on $^{16}$O and 
$^{56}$Fe. The shape followed by the RGF results fits well 
the slope shown by the data,  in particular approaching the peak region, 
where  the calculated cross sections are in reasonable agreement with the data. 
Although satisfactory on general grounds, the comparison with data 
cannot be conclusive until contributions beyond the QE peak, like meson 
exchange currents and $\Delta$ effects, which may
play a significant role in the analysis of data even at the maximum of the 
QE peak, are carefully evaluated.

The results of the RGF and RMF models have been compared 
for the inclusive QE electron scattering in~\cite{confee} and for CCQE neutrino 
scattering in~\cite{confcc}. Both models can describe successfully the 
behavior of electron scattering data. There are, however, some differences 
which increase with the momentum transfer. 
The RMF model uses as input the real, strong, energy-independent,
relativistic mean field potential that reproduces the saturation properties of 
nuclear matter and of the ground state of the nuclei involved. As such, it 
includes only purely nucleonic contribution and does not incorporate any 
information from scattering reactions. 
In contrast, the RGF, which uses as input a complex energy-dependent 
phenomenological ROP, incorporates information from 
scattering reactions and takes into account not only direct one-nucleon 
emission, but all the allowed final states. 
The imaginary part of the ROP includes the overall effect of the inelastic 
channels, which give different contributions at different energies. The 
differences between the RGF and RMF results can therefore be ascribed to the 
inelastic contributions which are incorporated in the RGF but not in the RMF, 
such as, for instance, re-scattering processes of the nucleon in its way out of the 
nucleus, non-nucleonic $\Delta$ excitations, which may arise during nucleon 
propagation, as well as to some multinucleon processes. 
These contributions are not included explicitly in the model, but can be 
recovered by the imaginary part of the ROP. 
The comparison between the RGF and RMF results can therefore be useful 
to evaluate the relevance of inelastic contributions. 

In the comparison with data, we may expect that the RGF can give a better 
description of the experimental cross sections which receive significant
contributions from non-nucleonic excitations and multi-nucleon processes.
This is expected to be the case \cite{compmini, Tina10} of  MiniBooNE CCQE data. 
While in electron-scattering experiments the beam energy is known and the cross
sections are given as a function of the energy transfer, in neutrino experiments
the energy and momentum transfer are not known and calculations are carried out 
over the energy range which is relevant for the neutrino flux. 
The flux-average procedure can include contributions from different kinematic 
regions where the neutrino flux has significant strength and processes other
than direct one-nucleon emission can be important. Part of 
these contributions are recovered in the RGF model by the imaginary part of 
the ROP.

\begin{figure}[h]
\centering
\includegraphics[scale=0.4]{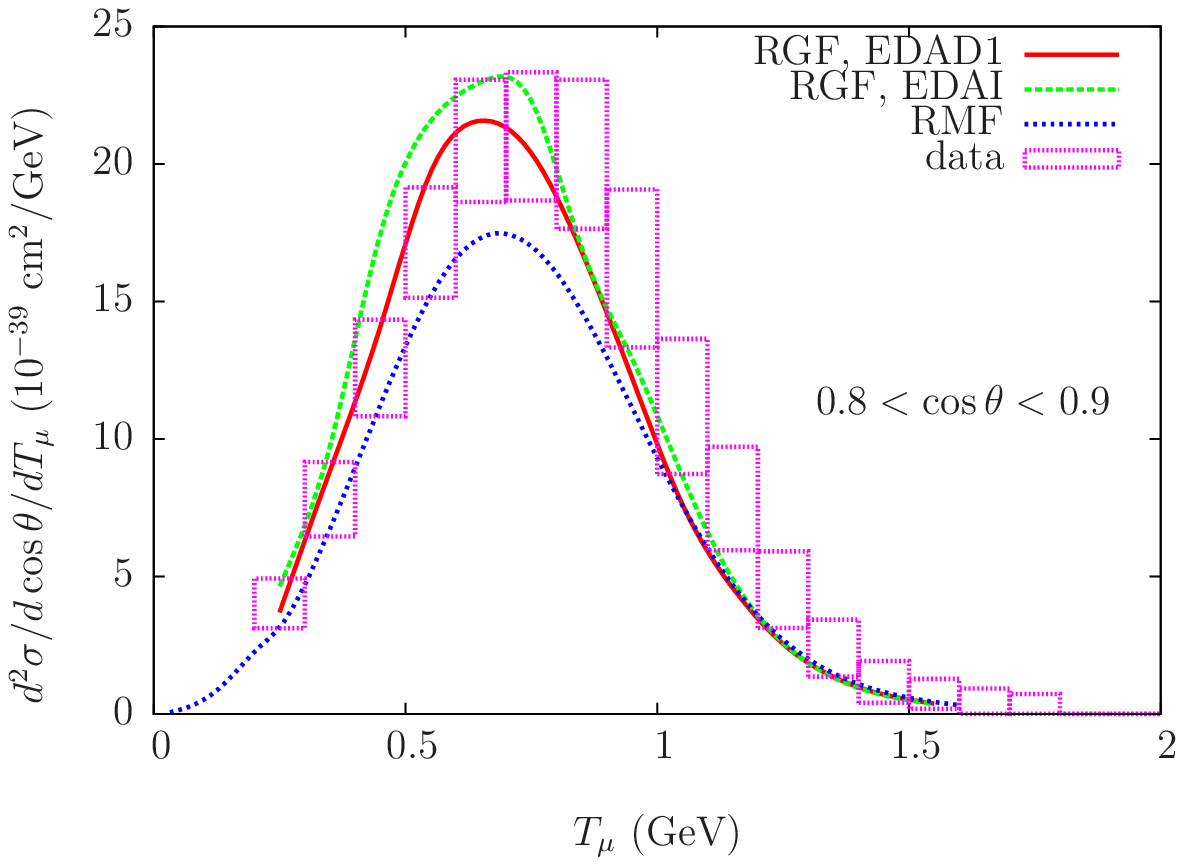}
\includegraphics[scale=0.4]{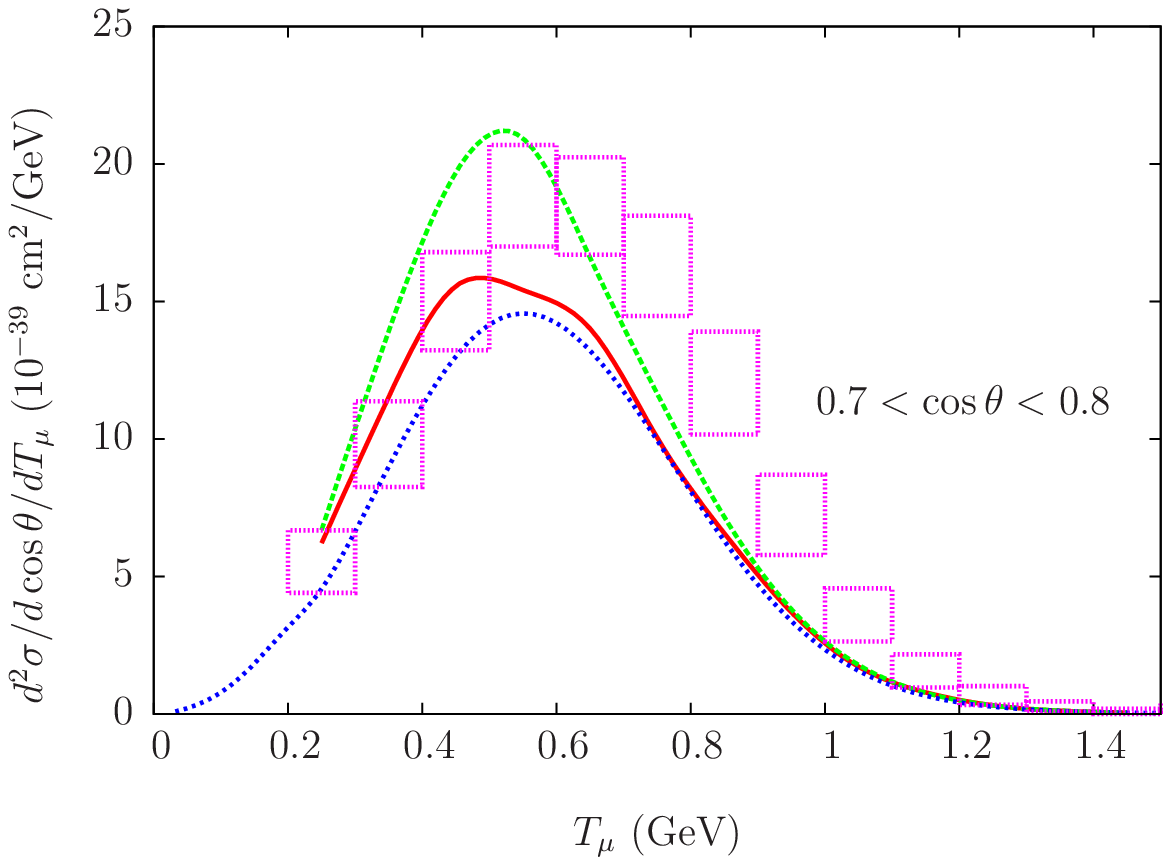}
\includegraphics[scale=0.4]{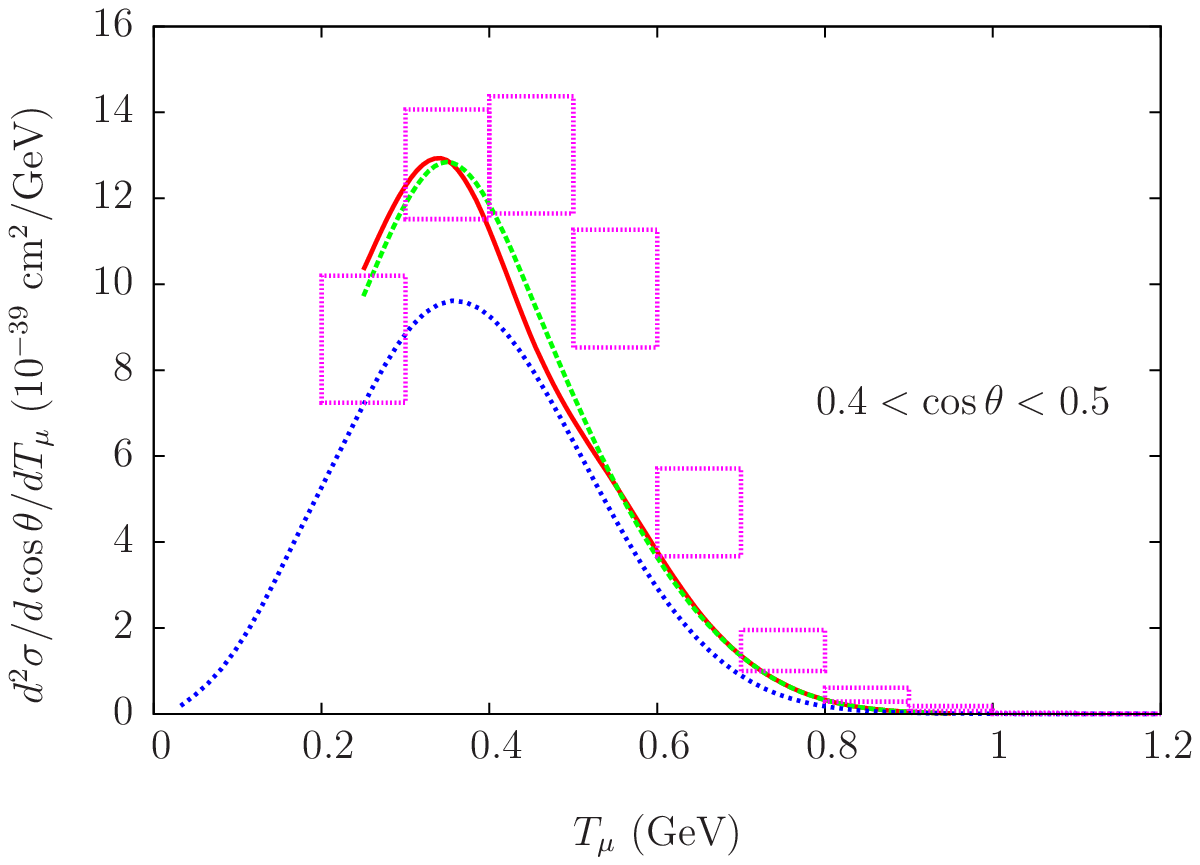}
\caption{Flux-averaged double-differential cross section per 
target nucleon for the CCQE $^{12}$C$(\nu_{\mu} , \mu ^-)$ reaction calculated 
in the RMF (blue line), the RGF-EDAD1 (red), and 
	RGF-EDAI (green), and displayed versus $T_\mu$ for various bins 
	of $\cos\theta$. In all the calculations the standard value of the 
	nucleon axial mass, {\it i.e.}, $M_A = 1.03$ GeV/$c^2$, has been used.
Experimental data from~\cite{miniboone}.
}
\label{MBCCQE}
\end{figure}

The CCQE $^{12}$C$(\nu_{\mu} , \mu ^-)$ cross sections  
measured by the MiniBooNE collaboration \cite{miniboone} have raised a strong 
debate over the role of the theoretical ingredients entering the description of 
the reaction. The experimental cross section is underestimated by the
relativistic Fermi gas (RFG) and by other more sophisticated models based
on the IA unless the nucleon axial mass  
$M_A$ is significantly enlarged with respect to the world average of 
all measured values (1.03 GeV/$c^2$ \cite{Bern02,bodek}), mostly obtained from 
deuteron data.
The larger axial mass obtained from the MiniBooNE data on carbon can also be 
interpreted as an effective way to include medium effects which are not taken 
into account by the RFG and by other models based on the IA.
Before drawing conclusions, the role of all nuclear effects must be carefully 
investigated. 

In Figure~\ref{MBCCQE} the flux-averaged 
RGF and RMF double-differential $^{12}$C$(\nu_{\mu},\mu^{-})$ cross sections 
are displayed as a function of the muon kinetic energy $T_\mu$ for 
various bins of $\cos\theta$, where $\theta$ is the muon scattering angle, 
and compared with the MiniBooNE data.
The shape followed by both RMF and RGF results fits well the slope shown 
by the data. The RMF cross sections generally underpredict the data, the RGF results
are generally larger than the RMF ones, in particular approaching the peak 
region, where the additional strength shown by the RGF produces cross sections 
in reasonable agreement with the data. The differences between the RGF-EDAI and RGF-EDAD1 results 
are in general of the order of the experimental errors.

\begin{figure}[h]
\centering
\includegraphics[scale=0.45]{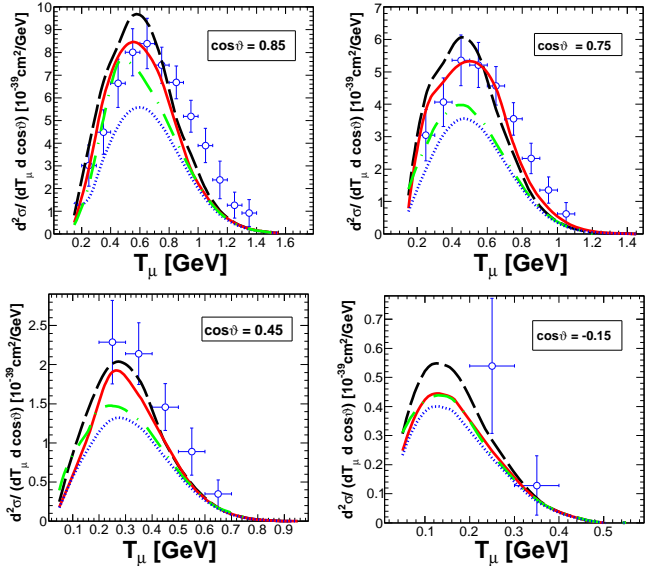}
\caption{Flux-averaged double-differential cross section per target nucleon for 
the CCQE $^{12}$C$\left(\bar\nu_{\mu}, \mu^+\right)$ reaction as a function of
$T_{\mu}$ 
for four angular bins of $\cos\theta$ calculated with the RGF-EDAD1 (solid lines) 
and the RGF-EDAI (dashed lines). The dotted lines are the rROP results calculated with 
the EDAI potential and the dot-dashed lines are the RPWIA results.
	Experimental data from~\cite{miniboonean}.
}
\label{MBCCQEAN}
\end{figure}

The flux-averaged double-differential $^{12}$C$\left(\bar\nu_{\mu}, \mu^+\right)$ cross 
sections are shown  in Figure~\ref{MBCCQEAN} as a function 
of $T_{\mu}$ for four angular bins of $\cos\theta$, ranging from forward to 
backward angles. The RGF results are compared with the RPWIA and rROP ones and 
with the MiniBooNE data~\cite{miniboonean}.
The rROP results are usually 15\% lower than the RPWIA ones, both rROP and 
RPWIA generally  underestimate the data. 
Larger cross sections, in better agreement with the data, are obtained with the 
RGF model, with both EDAD1 and EDAI. The differences between the RGF-EDAD1 and RGF-EDAI  
results are visible, although somewhat smaller than in the case of neutrino scattering.

\begin{figure}[h]
\centering
\includegraphics[scale=0.2,angle=-90]{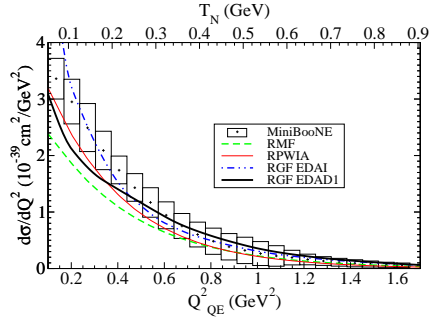}
\caption{NCE flux-averaged $(\nu N \rightarrow \nu N)$  cross section as
    a function of $Q^2$ calculated in the
    RPWIA (thin solid lines), RMF (dashed lines), RGF-EDAD1 (thick solid lines), and
    RGF-EDAI (dash-dotted lines). 
    Experimental data from~\cite{miniboonenc}.
}
\label{mininc}
\end{figure}
The MiniBooNE Collaboration has also measured \cite{miniboonenc}  the 
flux-averaged differential cross section as a function of 
the four-momentum transferred squared, $Q^2 = -q^{\mu}q_{\mu}$, for 
neutral-current elastic (NCE) neutrino scattering on CH$_2$ in a $Q^2$ range up 
to $\approx 1.65\ ($GeV/$c)^2$.
The analysis of $\nu$-nucleus NCE reactions introduces additional 
complications, as the final neutrino cannot be measured and a final hadron has 
to be detected: the cross sections are therefore semi-inclusive in the hadronic 
sector and inclusive in the leptonic one.
The description of semi-inclusive NCE scattering with the RGF approach can
recover important contributions that are not present in the RDWIA, 
which is appropriate for the exclusive scattering but neglects some final-state 
channels which can contribute to the semi-inclusive reaction. The RGF, however, 
describes the inclusive process and, as such, it may include channels which are not 
present in the semi-inclusive NCE 
measurements.  In comparison with the MiniBooNE NCE data, the RDWIA generally 
underpredicts the experimental cross section while the RGF gives a 
reasonable agreement with the NCE data \cite{prd}.

It is not easy to disentangle the role of contributions which may be
neglected in the RDWIA or spuriously added in the RGF, in particular
if we consider that both calculations make use of phenomenological ROP's. 
The comparison with the results of the RMF model, where 
only the purely nucleonic contribution is included, can be helpful for a deeper
understanding of FSI effects.

In Figure~\ref{MBCCQEAN} the RMF, RGF, and RPWIA cross sections for NCE 
$(\nu N \rightarrow \nu N)$ scattering are presented and compared with the
experimental data. The variable $Q^2_{QE}=2m_NT$ is defined assuming that the
target nucleon is at rest, $m_N$ being the nucleon mass and $T$ the total
kinetic energy of the outgoing nucleons.
The RMF result has a too soft $Q^2$ behavior to reproduce the data at small 
$Q^2$, while the RGF produces larger cross sections in better agreement with 
the data. The difference between the RGF-EDAD1 and RGF-EDAI results is
significant. 
The RGF-EDAI cross section is in accordance with the shape and the magnitude 
of the data, while the RGF-EDAD1 one lies below the data at the smallest 
values of $Q^2$ considered in the figure.
The RMF result gives the lowest cross section for low-to-intermediate 
values of $Q^2_{QE}$. 

\section{Conclusions}

A deep understanding of the reaction mechanism of neutrino-nucleus cross 
sections is mandatory for the determination of neutrino oscillation parameters. 
Reliable models are required where all nuclear effects are well 
under control. 
The RGF model has been discussed in this contribution.
This model was originally developed for QE electron scattering, it has been 
tested in comparison with electron-scattering data, and it has then been extended to 
neutrino-nucleus scattering. In the RGF model FSI are described in the 
inclusive scattering by the same complex ROP as in the exclusive scattering, but 
the imaginary part  redistributes and conserves the flux in all the channels. 
The RGF results are usually larger than the results of other models based 
on the RIA and can reproduce the CCQE MiniBooNE data without the 
need to increase the standard value of the nucleon axial mass. The enhancement 
of the RGF results is due to the 
translation to the inclusive strength of the overall effect of inelastic 
channels which are not incorporated in the RMF and in other models 
based on the IA. 
The use of phenomenological ROP's, however, does not allow us to 
disentangle different reaction processes and explain in detail the 
origin of the enhancement. 
The RGF results are also affected by uncertainties in
the determination of the phenomenological ROP. A better determination 
which closely fulfills the dispersion relations 
deserves further investigation. 

The relevance of contributions other than direct one-nucleon emission in 
kinematic regions where the experimental neutrino flux has significant
strength has been confirmed by different 
studies~\cite{benhar,Martini,Nieves,lala,ank}. 
Processes involving two-body currents, whose role is discussed 
in~\cite{amaro11b},  should also be taken into account explicitly and 
consistently in a model to clarify the role of multinucleon emission.

\section*{Acknowledgements}
We thank M.B. Barbaro, J.A. Caballero, F. Capuzzi, R. Gonz\'alez-Jim\'enez, M.V. 
Ivanov, F.D. Pacati,  and J.M. Ud\'{\i}as for the fruitful collaborations which 
led to the results reported in this contribution.

\end{document}